\documentclass[11pt]{article}
\usepackage{amssymb,amsmath,amsfonts}
\usepackage{graphicx}
\usepackage{graphics}
\usepackage{eepic,epsfig}

\begin{document}

\title{Casimir-Polder interaction between an atom and \\
a conducting wall in cosmic string spacetime}
\author{E. R. Bezerra de Mello$^{1}$\thanks{%
E-mail: emello@fisica.ufpb.br},\, V. B. Bezerra$^{1}$\thanks{%
E-mail: valdir@fisica.ufpb.br}, \, H. F. Mota$^{1,2}$, \, A. A. Saharian$%
^{3} $\thanks{%
E-mail: saharian@ysu.am} \\
\\
\textit{$^{1}$Departamento de F\'{\i}sica, Universidade Federal da Para\'{\i}%
ba}\\
\textit{58.059-970, Caixa Postal 5.008, Jo\~{a}o Pessoa, PB, Brazil}\vspace{%
0.3cm}\\
\textit{$^2$Department of Physics and Astronomy, Tufts University}\\
\textit{212 College Avenue, Medford, MA 02155, USA}\vspace{0.3cm}\\
\textit{$^3$Department of Physics, Yerevan State University}\\
\textit{Alex Manoogian Street, 0025 Yerevan, Armenia}}
\maketitle

\begin{abstract}
The Casimir-Polder interaction potential is evaluated for a polarizable
microparticle and a conducting wall in the geometry of a cosmic string
perpendicular to the wall. The general case of the anisotropic
polarizability tensor for the microparticle is considered. The corresponding
force is a function of the wall-microparticle and cosmic
string-microparticle distances. Depending on the orientation of the
polarizability tensor principal axes the force can be either attractive or
repulsive. The asymptotic behavior of the Casimir-Polder potential is
investigated at large and small separations compared to the wavelength of
the dominant atomic transitions. We show that the conical defect may be used
to control the strength and the sign of the Casimir-Polder force.
\end{abstract}

\bigskip

\section{Introduction}

Casimir-Polder (CP) interactions between atoms and surfaces are among the
most interesting manifestations of the electromagnetic quantum fluctuations
(for a review see \cite{Parseg}). The wide applications of the corresponding
forces in many areas of science and technology motivate the investigations
of various mechanisms to control their strength and sign. Recently large
efforts have been focused on the investigation of the nature of CP forces
and its dependence on the geometry of boundaries. In particular, there has
been extensive interest in geometries with repulsive CP forces (for a recent
discussion see \cite{Eber11} and references therein). In \cite{Saha12} we
have shown that the presence of topological defects can serve as an
additional tool for the control of the forces.

The formation of topological defects is predicted in some field-theoretical
and condensed matter systems as a result of symmetry breaking phase
transitions. In particular, within the framework of grand unified theories,
different types of topological defects could be produced in the early
universe \cite{Vile94,Cope10}. Among them, the cosmic strings have attracted
considerable attention. Although the recent observational data on the cosmic
microwave background radiation have ruled out cosmic strings as the primary
source for primordial density perturbations, they are still candidates for
the generation of a number of interesting astrophysical effects such as the
generation of gravitational waves, high-energy cosmic rays and gamma ray
bursts.

In the simplest model for a straight cosmic string the geometry outside the
string core is flat with a planar angle deficit. The corresponding
non-trivial topology leads to the modification of the zero-point
fluctuations of quantum fields. The change of vacuum properties due to this
modification has been discussed in a large number of papers (see, for
instance, the references given in \cite{Beze08}). The vacuum polarization
effect by a cosmic string in the background of de Sitter and anti-de Sitter
spacetimes is investigated in \cite{Beze09dS,Beze12AdS}. The distortion of
the electromagnetic field vacuum fluctuations by a cosmic string also gives
rise to a CP force acting on a polarizable microparticle. This force is
investigated in \cite{Bard10,Saha11EPJ}. It is shown that this force depends
on the eigenvalues for the polarizability tensor and on the orientation of
its principal axes. The CP force can be either repulsive or attractive with
respect to the string. For an isotropic polarizability tensor the force is
always repulsive. In \cite{Saha12} it was considered the influence of the
cosmic string on the CP force between a microparticle and a conducting
cylindrical shell coaxial with the string. The combined effects arising from
the topology of the cosmic string and boundaries on the vacuum energy and
stresses have been discussed in \cite{Beze08}, \cite{Brev95}-\cite{Nest11}.

In the present paper we investigate the influence of a conical defect
(cosmic string) on the CP force between a polarizable microparticle and a
conducting plate perpendicular to the defect. The paper is organized as
follows. In the next section we evaluate the retarded Green tensor for the
electromagnetic field in the geometry of a cosmic string with a conducting
plate. This tensor is used in section \ref{Sec:CP} to evaluate the CP force
acting on a polarizable microparticle for the general case of anisotropic
polarizability tensor. The CP force in the oscillator model for the
polarizability tensor is discussed in section \ref{Sec:Osc}. Section \ref%
{Sec:Conc} summarizes the main results.

\section{Retarded Green tensor}

\label{Sec:GF}

For a long straight cosmic string, at distances much larger than the core
radius, the corresponding spacetime geometry is flat with a planar angle
deficit $2\pi -\phi _{0}$. Considering the string situated along the $z$%
-axis, the line element, in cylindrical coordinates $(x^{1},x^{2},x^{3})=(r,%
\phi ,z)$, can be written as
\begin{equation}
ds^{2}=dt^{2}-dr^{2}-r^{2}d\phi ^{2}-dz^{2},  \label{eq1}
\end{equation}%
where $0\leqslant \phi \leqslant \phi _{0}=2\pi /q $, with $q$ being a
parameter associated with the cosmic string. This line element has been
derived in \cite{Vile81} in the weak-field and thin-string approximations.
In this case the angle deficit is related to the mass per unit length $\mu $
of the string by the formula $2\pi -\phi _{0}=\mu /m_{\mathrm{Pl}}^{2}$,
where $m_{\mathrm{Pl}}$ is the Planck mass. In the standard scenario for the
cosmic string formation in the early universe one has $\mu \sim \eta ^{2}$,
where $\eta $ is the energy scale of the phase transition at which the
string is formed. For GUT scale strings $\mu \ll m_{\mathrm{Pl}}^{2}$ and
the weak-field approximation is well justified. However, the validity of the
line element (\ref{eq1}) has been extended beyond the linear perturbation
theory by several authors \cite{Gott85} (see also \cite{Vile94}). In this
case the parameter $q$ need not to be close to 1. Note that the conical
defects appear as an effective geometry in a number of condensed matter
systems such as crystals, liquid crystals and quantum liquids (see, for
example, \cite{Nels02}).

Our main interest in the present paper is the CP force acting on a
polarizable microparticle (atom, molecule or any small object described by
an electric-dipole polarizability tensor) near a conducting plate
perpendicular to the cosmic string and located at $z=0$. For a microparticle
situated at a point $\mathbf{r}$, the CP interaction energy can be expressed
as
\begin{equation}
U(\mathbf{r})=\frac{1}{2\pi }\int_{0}^{\infty }d\xi \,\alpha _{jl}(i\xi
)G_{jl}^{(s)}(\mathbf{r},\mathbf{r};i\xi ),  \label{eq2}
\end{equation}%
where the summation is understood over $j,l=1,2,3$, $\alpha _{jl}(\omega )$
is the polarizability tensor of the particle and
\begin{equation}
G_{jl}^{(s)}(\mathbf{r},\mathbf{r}^{\prime };\omega )=\int_{-\infty
}^{+\infty }d\tau G_{jl}^{(s)}(x,x^{\prime })e^{i\omega \tau },  \label{eq3}
\end{equation}%
with $x=(t,\mathbf{r})$, $x^{\prime }=(t^{\prime },\mathbf{r}^{\prime })$, $%
\tau =t-t^{\prime }$. In (\ref{eq3}), $G_{jl}^{(s)}(x,x^{\prime })$ is given
by
\begin{equation}
G_{jl}^{(s)}(x,x^{\prime })=G_{jl}(x,x^{\prime })-G_{jl}^{\mathrm{(M)}%
}(x,x^{\prime }),  \label{Gsub}
\end{equation}%
where $G_{jl}(x,x^{\prime })$ is the retarded Green tensor for the
electromagnetic field in the geometry under consideration and $G_{jl}^{%
\mathrm{(M)}}(x,x^{\prime })$ is the retarded Green tensor in the
boundary-free Minkowski spacetime. The local geometry induced by the cosmic
string is flat and the subtracted Green tensor $G_{jl}^{(s)}(x,x^{\prime })$
is finite in the coincidence limit for points $z\neq 0$ and $r\neq 0$.

The retarded Green tensor for the electromagnetic field is given by the
expression
\begin{equation}
G_{jl}(x,x^{\prime })=-i\theta (\tau )\left\langle E_{j}(x)E_{l}(x^{\prime
})-E_{l}(x^{\prime })E_{j}(x)\right\rangle ,  \label{GT1}
\end{equation}%
where $\theta (x)$ is the unit-step function, $E_{j}(x)$ is the operator of
the $j$-th component of the electric field, and the angular brackets mean
the vacuum expectation value. If $\{E_{\alpha j}(x),E_{\alpha j}^{\ast
}(x)\} $ is a complete set of mode functions for the electric field, with
the collective index $\alpha $ specifying the modes, then the Green tensor
can be expressed as the sum over the modes:
\begin{equation}
G_{jl}(x,x^{\prime })=-i\theta (\tau )\sum_{\alpha }[E_{\alpha
j}(x)E_{\alpha l}^{\ast }(x^{\prime })-E_{\alpha l}(x^{\prime })E_{\alpha
j}^{\ast }(x)],  \label{GTms}
\end{equation}%
where the asterisk stands for complex conjugate. In (\ref{GTms}), $E_{\alpha
j}(x)$ is the $j$-th physical component of the electric field vector in
cylindrical coordinates and the values $j=1,2,3$ correspond to the $r,\phi
,z $ coordinates, respectively.

In the geometry under consideration we have two classes of mode functions
corresponding to the waves of the transverse magnetic (TM) and transverse
electric (TE) types. The corresponding mode functions for the electric
field, obeying the boundary condition $\mathbf{n}\times \mathbf{E}=0$ on the
conducting plate at $z=0$, with $\mathbf{n}$ being the normal to the plate,
are given by the expressions
\begin{equation}
E_{\alpha l}^{(\lambda )}(x)=\beta _{\alpha }E_{\alpha l}^{(\lambda
)}(r,z)e^{i\left( qm\phi -\omega t\right) },  \label{Emode}
\end{equation}%
where $\lambda =0,1$ correspond to TM and TE modes, respectively, and
\begin{equation}
\omega ^{2}=\gamma ^{2}+k^{2},\;m=0,\pm 1,\pm 2,\ldots ,  \label{omega}
\end{equation}%
with $0\leqslant \gamma <\infty $, $0\leqslant k<\infty $. The components of
the electric field in (\ref{Emode}), are given by
\begin{eqnarray}
E_{\alpha 1}^{(0)}(r,z) &=&-k\gamma J_{q|m|}^{\prime }(\gamma r)\sin (kz),
\notag \\
E_{\alpha 2}^{(0)}(r,z) &=&-ik\frac{qm}{r}J_{q|m|}(\gamma r)\sin (kz),
\label{ErzTM} \\
E_{\alpha 3}^{(0)}(r,z) &=&\gamma ^{2}J_{q|m|}(\gamma r)\cos (kz),  \notag
\end{eqnarray}%
for the TM modes and by the functions
\begin{eqnarray}
E_{\alpha 1}^{(1)}(r,z) &=&-\omega \frac{qm}{r}J_{q|m|}(\gamma r)\sin (kz),
\notag \\
E_{\alpha 2}^{(1)}(r,z) &=&-i\omega \gamma J_{q|m|}^{\prime }(\gamma r)\sin
(kz),  \label{ErzTE} \\
E_{\alpha 3}^{(1)}(r,z) &=&0.  \notag
\end{eqnarray}%
for the TE modes. In these expressions $J_{\nu }(x)$ is the Bessel function
and the prime means derivative with respect to the argument of the function.
As it is seen from the formulas for the mode functions, they are specified
by the set $\alpha =(\lambda ,\gamma ,m,k)$.

The problem under consideration is symmetric under the reflection with
respect to the plate, $z\rightarrow -z$. Here we consider the region $z>0$.
In this region the mode functions (\ref{Emode}) are normalized by the
condition
\begin{equation}
\int_{0}^{\infty }drr\int_{0}^{\phi _{0}}d\phi \int_{0}^{\infty }dz\,\mathbf{%
E}_{\alpha }^{(\lambda )}\cdot \mathbf{E}_{\alpha ^{\prime }}^{(\lambda
^{\prime })\ast }=2\pi \omega \delta _{\alpha \alpha ^{\prime }},  \label{NC}
\end{equation}%
where $\delta _{\alpha \alpha ^{\prime }}$ is understood as the Dirac delta
function for continuous components of the collective index $\alpha $ and as
the Kronecker delta for discrete ones. Substituting the expressions for the
mode functions, it can be seen that the normalization coefficient is given
by the expression
\begin{equation}
\beta _{\alpha }^{2}=\frac{2q}{\pi \gamma \omega },  \label{betalf}
\end{equation}%
for both TM and TE modes.

Substituting the mode functions into the mode sum (\ref{GTms}), for the
Green tensor one finds
\begin{eqnarray}
&& G_{jl}(x,x^{\prime }) =-2i\theta (\tau )\frac{q}{\pi }\sum_{m=-\infty
}^{+\infty }\sum_{\lambda =0,1}\int_{0}^{\infty }dk\int_{0}^{\infty }d\gamma
\frac{1}{\gamma \omega }  \notag \\
&& \qquad \times \left[ e^{iqm\Delta \phi -i\omega \tau }E_{\alpha
j}^{(\lambda )}(r,z)E_{\alpha l}^{(\lambda )\ast }(r^{\prime },z^{\prime
})-e^{-iqm\Delta \phi +i\omega \tau }E_{\alpha l}^{(\lambda )}(r^{\prime
},z^{\prime })E_{\alpha j}^{(\lambda )\ast }(r,z)\right] ,  \label{eq11}
\end{eqnarray}%
where $\Delta \phi =\phi -\phi ^{\prime }$. The spectral components of the
Green tensor are presented in the form:
\begin{eqnarray}
&&G_{jl}(\mathbf{r},\mathbf{r}^{\prime };i\xi )=-\frac{2q}{\pi }%
\sum_{m=-\infty }^{+\infty }\sum_{\lambda =0,1}\int_{0}^{\infty
}dk\int_{0}^{\infty }d\gamma \frac{1}{\gamma \omega }  \notag \\
&& \qquad \times \left[ E_{\alpha j}^{(\lambda )}(r,z)E_{\alpha l}^{(\lambda
)\ast }(r^{\prime },z^{\prime })\frac{e^{iqm\Delta \phi }}{\omega -i\xi }%
+E_{\alpha l}^{(\lambda )}(r^{\prime },z^{\prime })E_{\alpha j}^{(\lambda
)\ast }(r,z)\frac{e^{-iqm\Delta \phi }}{\omega +i\xi }\right] .  \label{eq12}
\end{eqnarray}

By taking into account the expressions (\ref{ErzTM}) and (\ref{ErzTE}), the
Green tensor may be decomposed as%
\begin{equation}
G_{jl}(\mathbf{r},\mathbf{r}^{\prime };i\xi )=G_{jl}^{(0)}(\mathbf{r},%
\mathbf{r}^{\prime };i\xi )+G_{jl}^{(b)}(\mathbf{r},\mathbf{r}^{\prime
};i\xi ),  \label{GTdec}
\end{equation}%
where $G_{jl}^{(0)}(\mathbf{r},\mathbf{r}^{\prime };i\xi )$ is the
corresponding function for the boundary-free cosmic string geometry and the
part $G_{jl}^{(b)}(\mathbf{r},\mathbf{r}^{\prime };i\xi )$ is induced by the
presence of the conducting plate at $z=0$. The CP interaction in the
boundary-free cosmic string geometry has been discussed in \cite%
{Bard10,Saha11EPJ} and here we will be mainly concerned with the
boundary-induced part.

The components of the tensor $G_{jl}^{(b)}(\mathbf{r},\mathbf{r}^{\prime
};i\xi )$ can be expressed in terms of the functions
\begin{eqnarray}
A(r,r^{\prime },\Delta \phi ,y,\xi) &=&\sideset{}{'}{\sum}_{m=0}^{\infty
}\cos (qm\Delta \phi )\int_{0}^{\infty }dk\cos (ky)  \notag \\
&&\times \int_{0}^{\infty }d\gamma \frac{\gamma J_{qm}(\gamma
r)J_{qm}(\gamma r^{\prime })}{\omega ^{2}+\xi ^{2}},  \label{eq15}
\end{eqnarray}%
and%
\begin{eqnarray}
B(r,r^{\prime },\Delta \phi ,y,\xi) &=&\sum_{j=\pm 1}\sideset{}{'}{\sum}%
_{m=0}^{\infty }\cos (qm\Delta \phi )\int_{0}^{\infty }dk\cos (ky)  \notag \\
&&\times \int_{0}^{\infty }d\gamma \frac{\gamma J_{qm-j}(\gamma
r)J_{qm-j}(\gamma r^{\prime })}{\omega ^{2}+\xi ^{2}},  \label{eq16}
\end{eqnarray}%
where the prime on the sign of the sum means that the term $m=0$ should be
taken with weight $1/2$. For the diagonal components one has the expressions
\begin{eqnarray}
G_{11}^{(b)}(\mathbf{r},\mathbf{r}^{\prime };i\xi ) &=&-\frac{2q}{\pi }\left[
\partial _{z}^{2}B(r,r^{\prime },\Delta \phi ,z+z^{\prime })+\frac{2}{%
rr^{\prime }}\partial _{\Delta \phi }^{2}A(r,r^{\prime },\Delta \phi
,z+z^{\prime },\xi)\right] ,  \notag \\
G_{22}^{(b)}(\mathbf{r},\mathbf{r}^{\prime };i\xi ) &=&-\frac{2q}{\pi }\left[
\partial _{z}^{2}B(r,r^{\prime },\Delta \phi ,z+z^{\prime })-2\partial
_{r}\partial _{r^{\prime }}A(r,r^{\prime },\Delta \phi ,z+z^{\prime },\xi)%
\right] ,  \label{G22b} \\
G_{33}^{(b)}(\mathbf{r},\mathbf{r}^{\prime };i\xi ) &=&\frac{4q}{\pi }\left(
-\partial _{z}^{2}+\xi ^{2}\right) A(r,r^{\prime },\Delta \phi ,z+z^{\prime
},\xi).  \notag
\end{eqnarray}%
The off-diagonal components are presented as
\begin{eqnarray}
G_{12}^{(b)}(\mathbf{r},\mathbf{r}^{\prime };i\xi ) &=&-\frac{4q}{\pi
rr^{\prime }}\left[ 2\partial _{\xi ^{2}}\partial _{z}^{2}\partial _{\Delta
\phi }+r^{\prime }\partial _{r^{\prime }}\partial _{\Delta \phi }\right]
A(r,r^{\prime },\Delta \phi ,z+z^{\prime },\xi),  \notag \\
G_{13}^{(b)}(\mathbf{r},\mathbf{r}^{\prime };i\xi ) &=&-\frac{4q}{\pi }%
\partial _{z}\partial _{r}A(r,r^{\prime },\Delta \phi ,z+z^{\prime },\xi),
\label{G13b} \\
G_{23}^{(b)}(\mathbf{r},\mathbf{r}^{\prime };i\xi ) &=&-\frac{4q}{\pi r}%
\partial _{\Delta \phi }\partial _{z}A(r,r^{\prime },\Delta \phi
,z+z^{\prime },\xi).  \notag
\end{eqnarray}%
The remained off-diagonal components of the Green tensor are obtained from
those in (\ref{G13b}) by using the relation
\begin{equation}
G_{lj}(\mathbf{r}^{\prime },\mathbf{r};-i\xi )=G_{jl}(\mathbf{r},\mathbf{r}%
^{\prime };i\xi ).  \label{Gjl}
\end{equation}

With formulas (\ref{G22b}) and (\ref{G13b}), the evaluation of the Green
tensor is reduced to the evaluation of the functions (\ref{eq15}) and (\ref%
{eq16}). For the function (\ref{eq15}) one has the following representation
\cite{Saha11EPJ}:%
\begin{equation}
A(r,r^{\prime },\Delta \phi ,z+z^{\prime },\xi )=\frac{\pi }{4q}\bigg[%
\sum_{k}\frac{e^{-\xi u_{k}}}{u_{k}}-\frac{q}{2\pi }\sum_{j=\pm
1}\int_{0}^{\infty }dx\frac{\sin (q\pi +jq\Delta \phi )e^{-\xi v(x)}/v(x)}{%
\cosh (qx)-\cos (q\pi +jq\Delta \phi )}\bigg],  \label{A2}
\end{equation}%
where we have defined%
\begin{eqnarray}
u_{k} &=&\sqrt{r^{2}+r^{\prime 2}+(z+z^{\prime })^{2}-2rr^{\prime }\cos
(2\pi k/q-\Delta \phi )},  \notag \\
v(x) &=&\sqrt{r^{2}+r^{\prime 2}+(z+z^{\prime })^{2}+2rr^{\prime }\cosh x}.
\label{vx}
\end{eqnarray}%
In the first term on the right-hand side of (\ref{A2}) the summation goes
under the condition%
\begin{equation}
-q/2+q\Delta \phi /(2\pi )\leqslant k\leqslant q/2+q\Delta \phi /(2\pi ).
\label{SumCond}
\end{equation}%
A similar representation takes place for the function (\ref{eq16}) \cite%
{Saha11EPJ}:%
\begin{eqnarray}
B(r,r^{\prime },\Delta \phi ,z+z^{\prime },\xi ) &=&\frac{2r}{r^{\prime }}%
A(r,r^{\prime },\Delta \phi ,z+z^{\prime },\xi )+\frac{\pi }{2q\xi }\frac{1}{%
r^{\prime }}\partial _{r}\bigg[\sum_{k}e^{-\xi u_{k}}  \notag \\
&&-\frac{q}{2\pi }\sum_{j=\pm 1}\int_{0}^{\infty }dx\frac{\sin (q\pi
+jq\Delta \phi )e^{-\xi v(x)}}{\cosh (qx)-\cos (q\pi +jq\Delta \phi )}\bigg].
\label{B2}
\end{eqnarray}

For the evaluation of the CP potential, we need the components of the
boundary-induced part of the Green tensor $G_{jl}^{(b)}(\mathbf{r},\mathbf{r}%
^{\prime };i\xi )$ in the coincidence limit: $\mathbf{r}^{\prime
}\rightarrow \mathbf{r}$. In order to evaluate $\partial _{\Delta \phi
}^{2}A(r,r^{\prime },\Delta \phi ,z+z^{\prime },\xi)$, in the coincidence
limit, it is convenient to use the relation
\begin{equation*}
\lim_{\mathbf{r}^{\prime }\rightarrow \mathbf{r}}\partial _{\Delta \phi
}^{2}A(r,r^{\prime },\Delta \phi ,\Delta ^{\prime }z,\xi)=-\lim_{\mathbf{r}%
^{\prime }\rightarrow \mathbf{r}}[r\partial _{r}(r\partial
_{r})-4r^{2}\partial _{z^{2}}(r\partial _{r}+1)]A(r,r^{\prime },\Delta \phi
,z+z^{\prime },\xi).
\end{equation*}%
For the diagonal components one finds the following expression:
\begin{eqnarray}
G_{ll}^{(b)}(\mathbf{r},\mathbf{r};i\xi ) &=&-2\xi ^{3}\bigg[%
\sideset{}{'}{\sum}_{k=0}^{[q/2]}f_{l}(2\xi \sqrt{r^{2}s_{k}^{2}+z^{2}}%
,s_{k},z)-\frac{q}{\pi }\sin (q\pi )  \notag \\
&&\times \int_{0}^{\infty }dy\frac{f_{l}(2\xi \sqrt{r^{2}\cosh ^{2}y+z^{2}}%
,\cosh y,z)}{\cosh (2qy)-\cos (q\pi )}\bigg],  \label{Gbll}
\end{eqnarray}%
where $[q/2]$ means the integer part of $q/2$ and we have introduced the
notation%
\begin{equation}
s_{k}=\sin (\pi k/q).  \label{sk}
\end{equation}%
As before, the prime on the sign of sum in (\ref{Gbll}) means that the term $%
k=0$ is taken with weight 1/2. In (\ref{Gbll}) we have defined the function
\begin{equation}
f_{l}(u,v,z)=e^{-u}\sum_{p=1}^{3}[b_{lp}(v)u^{p-4}+4z^{2}\xi
^{2}c_{lp}(v)u^{p-6}],  \label{fl}
\end{equation}%
with
\begin{eqnarray}
b_{lp}(v) &=&b_{lp}^{(0)}+b_{lp}^{(1)}v^{2},  \notag \\
c_{lp}(v) &=&c_{lp}^{(0)}+c_{lp}^{(1)}v^{2}.  \label{eq22}
\end{eqnarray}%
The coefficients in (\ref{eq22}) are given by the matrices
\begin{equation}
b_{lp}^{(0)}=\left(
\begin{array}{ccc}
1 & 1 & 1 \\
-2 & -2 & 0 \\
-1 & -1 & -1%
\end{array}%
\right) ,\;\;\;\ b_{lp}^{(1)}=\left(
\begin{array}{ccc}
1 & 1 & -1 \\
1 & 1 & -1 \\
0 & 0 & 0%
\end{array}%
\right) ,  \label{blp}
\end{equation}%
and%
\begin{equation}
c_{lp}^{(0)}=\left(
\begin{array}{ccc}
0 & 0 & 0 \\
3 & 3 & 1 \\
3 & 3 & 1%
\end{array}%
\right) ,\;\;\;\ c_{lp}^{(1)}=\left(
\begin{array}{ccc}
-3 & -3 & -1 \\
-3 & -3 & -1 \\
0 & 0 & 0%
\end{array}%
\right) ,  \label{clp}
\end{equation}
where the rows and columns are numbered by $l$ and $p$, respectively.

In the coincidence limit $\mathbf{r}^{\prime }\rightarrow \mathbf{r}$, the
only nonzero off-diagonal component of the boundary-induced part of the
Green tensor is $G_{13}^{(b)}(\mathbf{r},\mathbf{r};i\xi )$. From (\ref{G13b}%
) one finds that this component is given by%
\begin{eqnarray}
G_{13}^{(b)}(\mathbf{r},\mathbf{r};i\xi ) &=&-8rz\xi ^{5}\bigg[%
\sideset{}{'}{\sum}_{k=0}^{[q/2]}s_{k}^{2}f_{13}(2\xi \sqrt{%
r^{2}s_{k}^{2}+z^{2}})-\frac{q}{\pi }\sin (q\pi )  \notag \\
&&\times \int_{0}^{\infty }dy\frac{f_{13}(2\xi \sqrt{r^{2}\cosh ^{2}y+z^{2}})%
}{\cosh (2qy)-\cos (q\pi )}\cosh ^{2}y\bigg],  \label{G13bc}
\end{eqnarray}%
with the notation%
\begin{equation}
f_{13}(u)=u^{-5}e^{-u}(u^{2}+3u+3).  \label{f13}
\end{equation}%
Note that the problem under consideration has less symmetry than the one for
a conducting cylindrical boundary coaxial with the string, considered in
\cite{Saha12}, and in the coincidence limit the Green tensor is non-diagonal.

For integer values of the parameter $q$, formulas (\ref{Gbll}) and (\ref%
{G13bc}) are reduced to
\begin{eqnarray}
G_{ll}^{(b)}(\mathbf{r},\mathbf{r};i\xi ) &=&-\xi
^{3}\sum_{k=0}^{q-1}f_{l}(2\xi \sqrt{r^{2}s_{k}^{2}+z^{2}},s_{k},z),  \notag
\\
G_{13}^{(b)}(\mathbf{r},\mathbf{r};i\xi ) &=&-4rz\xi
^{5}\sum_{k=0}^{q-1}s_{k}^{2}f_{13}(2\xi \sqrt{r^{2}s_{k}^{2}+z^{2}}).
\label{Gllbint}
\end{eqnarray}%
The $k=0$ terms in these expressions correspond to the boundary-induced part
of the Green tensor for a conducting plate in Minkowski spacetime. Note that
the corresponding off-diagonal component vanishes.

\section{Casimir-Polder potential}

\label{Sec:CP}

Having the components of the retarded Green tensor, we can evaluate the CP
potential using formula (\ref{eq2}). Taking into account Eq. (\ref{GTdec}),
the potential may be decomposed as%
\begin{equation}
U(\mathbf{r})=U_{0}(\mathbf{r})+U_{b}(\mathbf{r}),  \label{Udec}
\end{equation}%
where
\begin{equation}
U_{0}(\mathbf{r})=\frac{1}{2\pi }\int_{0}^{\infty }d\xi \,\alpha _{jl}(i\xi
)[G_{jl}^{(0)}(\mathbf{r},\mathbf{r};i\xi )-G_{jl}^{\mathrm{(M)}}(\mathbf{r},%
\mathbf{r};i\xi )]  \label{U0}
\end{equation}%
is the potential in a boundary-free cosmic string geometry and the part%
\begin{equation}
U_{b}(\mathbf{r})=\frac{1}{2\pi }\int_{0}^{\infty }d\xi \,\alpha _{jl}(i\xi
)G_{jl}^{(b)}(\mathbf{r},\mathbf{r};i\xi )  \label{Ub}
\end{equation}%
is induced by the plate at $z=0$.

Substituting expressions (\ref{Gbll}) and (\ref{G13bc}), for the components
of the Green tensor, into (\ref{Ub}), we get the following result
\begin{equation}
U_{b}(\mathbf{r})=-\frac{1}{16\pi }\left[ \sideset{}{'}{\sum}%
_{k=0}^{[q/2]}f(r,z,s_{k})-\frac{q}{\pi }\int_{0}^{\infty }dy\frac{\sin
(q\pi )f(r,z,\cosh y)}{\cosh (2qy)-\cos (q\pi )}\right] .  \label{Ub1}
\end{equation}%
Here we have introduced the notation%
\begin{equation}
f(r,z,x)=\sum_{l,p=1}^{3}\left[ (r^{2}x^{2}+z^{2})b_{lp}(x)+z^{2}c_{lp}(x)%
\right] \frac{h_{lp}(2\sqrt{r^{2}x^{2}+z^{2}})}{(r^{2}x^{2}+z^{2})^{3}}%
+2rzx^{2}\frac{h(2\sqrt{r^{2}x^{2}+z^{2}})}{(r^{2}x^{2}+z^{2})^{3}},
\label{frzx}
\end{equation}%
with the functions%
\begin{eqnarray}
h_{lp}(y) &=&\int_{0}^{\infty }du\,u^{p-1}e^{-u}\alpha _{ll}(iu/y),  \notag
\\
h(y) &=&\int_{0}^{\infty }du\,e^{-u}(u^{2}+3u+3)\alpha _{13}(iu/y).
\label{hlp}
\end{eqnarray}%
Assuming that $r\gg z$, the dominant contribution to the CP potential comes
from the $k=0$ term and, to the leading order, the potential coincides with
the corresponding potential for a plate in Minkowski spacetime: $U_{b}(%
\mathbf{r})\approx U_{b}^{\mathrm{(M)}}(\mathbf{r})$ (see Eq. (\ref{UbM})
below). In the opposite limit, when $r\ll z$, the potential is dominated by
the pure string part $U_{0}(\mathbf{r})$.

\begin{figure}[tbph]
\begin{center}
\epsfig{figure=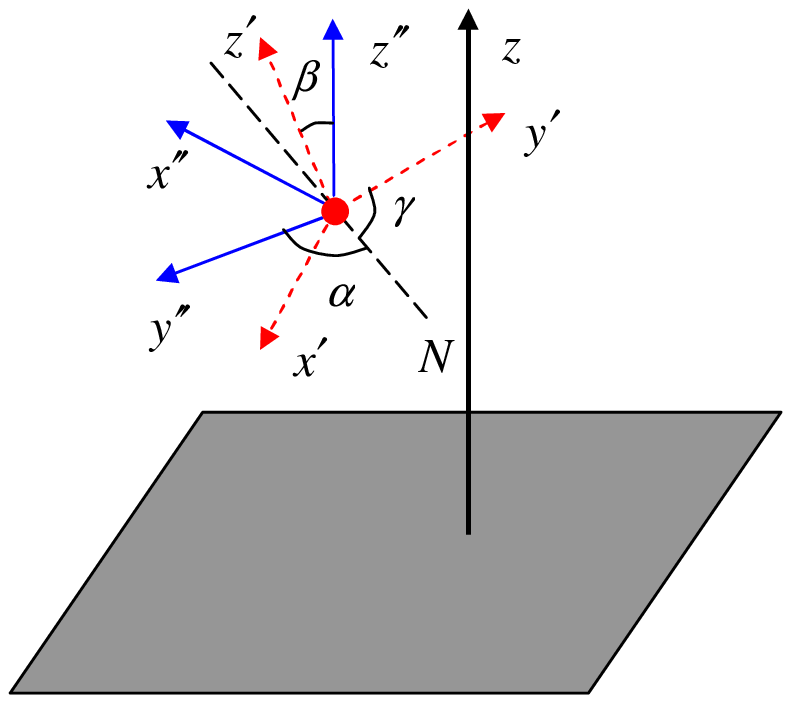,width=7.cm,height=6.cm}
\end{center}
\caption{Microparticle near a conducting plate. The cosmic string is
directed along the $z$-axis.}
\label{fig1}
\end{figure}

In (\ref{hlp}), $\alpha _{jl}(i\xi )$ are the physical components of the
polarizability tensor in the cylindrical coordinates corresponding to line
element (\ref{eq1}). These components depend on the orientation of the
polarizability tensor principal axes. As a consequence, the CP potential
depends on the distance of the microparticle from the string, on the
distance from the plate and on the angles determining the orientation of the
principal axes. Let $x^{\prime l}=(x^{\prime },y^{\prime },z^{\prime })$ be
the Cartesian coordinates with the origin at the location of the
microparticle and with the axes directed along the principal axes of the
polarizability tensor (see Figure \ref{fig1}). We also introduce the
intermediate Cartesian coordinates $x^{\prime \prime l}=(x^{\prime \prime
},y^{\prime \prime },z^{\prime \prime })$ with the same origin and with the $%
z^{\prime \prime }$ axis parallel to the string and with the string
coordinate $x^{\prime \prime }=-r$. Let $\beta _{ln}$ be the cosine of the
angle between the axes $x^{\prime \prime l}$ and $x^{\prime n}$. One has $%
\sum_{n=1}^{3}\beta _{ln}^{2}=1$. The coefficients $\beta _{ln}$ can be
expressed in terms of the Euler angles $(\alpha ,\beta ,\gamma )$ (see
Figure \ref{fig1}) determining the orientation of the principal axes with
respect to the coordinate system $x^{\prime \prime l}$ (see, for example,
\cite{Korn68}). The corresponding matrix $\hat{R}$, with the elements $\beta
_{ln}$, is given by the expression%
\begin{equation}
\hat{R}=\left(
\begin{array}{rrr}
\cos \alpha \cos \beta \cos \gamma -\sin \alpha \sin \gamma  & -\cos \alpha
\cos \beta \sin \gamma -\sin \alpha \cos \gamma  & \cos \alpha \sin \beta
\\
\sin \alpha \cos \beta \cos \gamma +\cos \alpha \sin \gamma  & -\sin \alpha
\cos \beta \sin \gamma +\cos \alpha \cos \gamma  & \sin \alpha \sin \beta
\\
-\sin \beta \cos \gamma  & \sin \beta \sin \gamma  & \cos \beta
\end{array}%
\right) ,  \label{betmatrix}
\end{equation}%
where $\beta $ is the angle between the axes $z^{\prime }$ and $z^{\prime
\prime }$, $\gamma $ ($\alpha $) is the angle between the axis $y^{\prime }$
($y^{\prime \prime }$) and the line of nodes (line of the intersection of
the $x^{\prime }y^{\prime }$ and the $x^{\prime \prime }y^{\prime \prime }$
coordinate planes, the line $N$ in Figure \ref{fig1}). For the diagonal
components of the polarizability tensor appearing in (\ref{hlp}) we have%
\begin{equation}
\alpha _{ll}(\omega )=\sum_{n=1}^{3}\beta _{ln}^{2}\alpha _{n}(\omega ),
\label{alfllom}
\end{equation}%
where $\alpha _{n}(\omega )$ are the principal values of the polarizability
tensor. The off-diagonal component can be written as $\alpha _{13}(\omega
)=\sum_{n=1}^{3}\beta _{1n}\beta _{3n}\alpha _{n}(\omega )$, or by taking
into account (\ref{betmatrix}):
\begin{eqnarray}
\alpha _{13}(\omega ) &=&\sin \beta \lbrack \left( \alpha _{1}(\omega
)-\alpha _{2}(\omega )\right) \sin \gamma \left( \sin \alpha \cos \gamma
+\cos \alpha \sin \gamma \cos \beta \right)   \notag \\
&&+\left( \alpha _{3}(\omega )-\alpha _{1}(\omega )\right) \cos \alpha \cos
\beta ].  \label{alf13}
\end{eqnarray}%
In the isotropic case $\alpha _{n}(\omega )\equiv \alpha (\omega )$ and we
have $\alpha _{jl}(\omega )=\alpha (\omega )\delta _{jl}$. When%
\begin{equation}
\alpha _{1}(\omega )=\alpha _{2}(\omega ),  \label{alfeq}
\end{equation}
from the general expressions one has simpler relations%
\begin{eqnarray}
\alpha _{ll}(\omega ) &=&\alpha _{1}(\omega )+\left[ \alpha _{3}(\omega
)-\alpha _{1}(\omega )\right] \beta _{l3}^{2},  \notag \\
\alpha _{13}(\omega ) &=&\frac{1}{2}\left[ \alpha _{3}(\omega )-\alpha
_{1}(\omega )\right] \cos \alpha \sin (2\beta ).  \label{Alfspec}
\end{eqnarray}%
In this special case the CP potential does not depend on the angle $\gamma $.

For integer values of the parameter $q$, the general formula (\ref{Ub1}) is
further simplified to
\begin{equation}
U_{b}(\mathbf{r})=-\frac{1}{32\pi }\sum_{k=0}^{q-1}f(r,z,s_{k}).
\label{Ubqint}
\end{equation}%
The $k=0$ term in this expression (and also in (\ref{Ub1})) coincides with
the CP potential for the geometry of a plate in Minkowski spacetime, $U_{b}^{%
\mathrm{(M)}}(\mathbf{r})$. Taking into account the expression (\ref{frzx}),
we find
\begin{equation}
U_{b}^{\mathrm{(M)}}(\mathbf{r})=-\frac{z^{-4}}{32\pi }\int_{0}^{\infty
}dxe^{-x}[\left( 1+x+x^{2}\right) \sum_{n=1}^{3}\alpha _{n}(ix/2z)+\left(
1+x-x^{2}\right) \alpha _{33}(ix/2z)],  \label{UbM}
\end{equation}%
where%
\begin{equation}
\alpha _{33}(\omega )=[\alpha _{1}(\omega )\cos ^{2}\gamma +\alpha
_{2}(\omega )\sin ^{2}\gamma ]\sin ^{2}\beta +\alpha _{3}(\omega )\cos
^{2}\beta .  \label{alf33}
\end{equation}%
For $\alpha _{n}(ix/2z)>0$, the corresponding CP force is always attractive.
This force does not depend on the angle $\alpha $. In the special case (\ref%
{alfeq}), we have no dependence on the angle $\gamma $ as well.

Now let us consider the asymptotic of the CP potential (\ref{Ub1}) at large
distances from the string and from the boundary compared to the wavelength
of the main atomic absorption lines. In this case, the expression for the
function $f(r,z,x)$ takes the form
\begin{equation}
f(r,z,x)\approx 4\sum_{l=1}^{3}\alpha _{ll}(0)\frac{\left(
b_{l}r^{2}+c_{l}z^{2}\right) x^{2}+z^{2}}{(r^{2}x^{2}+z^{2})^{3}}+\frac{%
16rzx^{2}\alpha _{13}(0)}{(r^{2}x^{2}+z^{2})^{3}},  \label{flarge}
\end{equation}%
with the coefficients%
\begin{equation}
b_{l}=(1,-1,-1),\;c_{l}=(-2,-2,0).  \label{albl}
\end{equation}%
If, in addition, $z\gg r$, the total CP potential (\ref{Udec}) is dominated
by the pure string part $U_{0}(\mathbf{r})$ and to the leading order one
gets
\begin{equation}
U(\mathbf{r})\approx U_{0}(\mathbf{r})\approx \frac{\left( q^{2}-1\right)
\left( q^{2}+11\right) }{360\pi r^{4}}\left[ \alpha _{11}(0)-\alpha
_{22}(0)+\alpha _{33}(0)\right] .  \label{U0large}
\end{equation}%
In the special case (\ref{alfeq}) the explicit dependence of the potential
on the orientation of the principal axes is given by taking into account the
relation%
\begin{equation}
\alpha _{11}(0)-\alpha _{22}(0)+\alpha _{33}(0)=\alpha _{3}(0)+2\left[
\alpha _{1}(0)-\alpha _{3}(0)\right] \sin ^{2}\alpha \sin ^{2}\beta .
\label{alfcomb}
\end{equation}%
In this case, for a fixed value of $r$, the equilibrium
orientation corresponds to $\alpha =\beta =0$ for $\alpha
_{1}(0)>\alpha _{3}(0)$ and to $\alpha =\beta =\pi /2$ for $\alpha
_{1}(0)<\alpha _{3}(0)$.

The $z$-projection of the force is determined by the boundary-induced part
of the CP. At distances from the boundary much larger than the relevant
transition wavelengths, assuming $z\gg r$, the function $f(r,z,x)$ given by (%
\ref{flarge}) can be written, approximately, as
\begin{equation}
f(r,z,x)\approx \frac{4}{z^{4}}\sum_{m=1}^{3}\alpha _{m}(0)\left[
1-2x^{2}\left( 1-\beta _{3m}^{2}\right) \right] .  \label{frzAs}
\end{equation}%
The corresponding force is attractive with respect to the plate. In
particular, for integer $q\geqslant 2$, after the summation over $k$ in (\ref%
{Ubqint}), one finds%
\begin{equation}
U_{b}(\mathbf{r})\approx -\frac{q}{8\pi z^{4}}\alpha _{33}(0).
\label{UbAsInt}
\end{equation}%
Note from (\ref{UbM}) that for a plate in Minkowski spacetime, at large
distances, one has
\begin{equation}
U_{b}^{\mathrm{(M)}}(\mathbf{r})\approx -\frac{1}{8\pi z^{4}}%
\sum_{m=1}^{3}\alpha _{m}(0).  \label{UbmAs}
\end{equation}%
The latter does not depend on the orientation of the principal axes for the
polarizability tensor. Comparing with (\ref{UbAsInt}), we see that this is
not the case when the string is present.

At distances smaller than the wavelength of the main atomic absorption lines
the dominant contribution to the CP potential comes from the $p=1$ term and
from the last term in (\ref{frzx}). Up to the leading order, the CP
potential is given by (\ref{Ub1}) with%
\begin{eqnarray}
f(r,z,x) &\approx &2\sum_{l}^{3}\frac{%
(r^{2}x^{2}+z^{2})b_{l1}(x)+z^{2}c_{l1}(x)}{(r^{2}x^{2}+z^{2})^{5/2}}%
\int_{0}^{\infty }du\,\alpha _{ll}(iu)  \notag \\
&&+\frac{12rzx^{2}}{(r^{2}x^{2}+z^{2})^{5/2}}\int_{0}^{\infty }du\,\alpha
_{13}(iu).  \label{frzxsm}
\end{eqnarray}%
If in addition, $r\ll z$, one finds%
\begin{equation}
f(r,z,x)\approx \frac{2}{z^{3}}[\left( 1-2x^{2}\right)
\sum_{l=1}^{3}\int_{0}^{\infty }du\,\alpha _{l}(iu)+\left( 1+2x^{2}\right)
\int_{0}^{\infty }du\,\alpha _{33}(iu)],  \label{frzxsmr}
\end{equation}%
where $\alpha _{33}(\omega )$ is given by the expression (\ref{alf33}). For
integer values $q\geqslant 2$, the CP potential has the asymptotic form%
\begin{equation}
U_{b}(\mathbf{r})\approx -\frac{q}{8\pi z^{3}}\int_{0}^{\infty }du\,\alpha
_{33}(iu).  \label{Ubqintsm}
\end{equation}%
Note that for the CP potential in the Minkowski spacetime at small distances
we have%
\begin{equation}
U_{b}^{\mathrm{(M)}}(\mathbf{r})\approx -\frac{z^{-3}}{16\pi }%
\int_{0}^{\infty }du\,[\sum_{l=1}^{3}\alpha _{l}(iu)+\alpha _{33}(iu)].
\label{UbMsm}
\end{equation}

In the isotropic case one has $\alpha _{jl}(\omega )=\alpha (\omega )\delta
_{jl}$ and the expression for the function (\ref{frzx}) takes the form
\begin{eqnarray}
&& f(r,z,x) =2(r^{2}x^{2}+z^{2})^{-3}\int_{0}^{\infty }due^{-u}\alpha (iu/2%
\sqrt{r^{2}x^{2}+z^{2}})  \notag \\
&& \qquad \times \left\{ (x^{2}-1)(1+u)\left(
r^{2}x^{2}-2z^{2}\right) +u^{2}\left[
z^{2}(1-2x^{2})-r^{2}x^{4}\right] \right\} . \label{frzxiz}
\end{eqnarray}%
At large distances we find%
\begin{equation}
U_{b}(\mathbf{r})\approx \frac{\alpha (0)}{4\pi z^{4}}\left[ %
\sideset{}{'}{\sum}_{k=0}^{[q/2]}g_{1}(r/z,s_{k})-\frac{q}{\pi }%
\int_{0}^{\infty }dy\frac{\sin (q\pi )g_{1}(r/z,\cosh y)}{\cosh (2qy)-\cos
(q\pi )}\right] ,  \label{Ubrizlarge}
\end{equation}%
where%
\begin{equation}
g_{1}(y,x)=\frac{y^{2}x^{2}+4x^{2}-3}{(y^{2}x^{2}+1)^{3}}.  \label{grx}
\end{equation}%
In particular, for integer values of $q$ one has%
\begin{equation}
U_{b}(\mathbf{r})\approx \frac{\alpha (0)}{8\pi }\sum_{k=0}^{q-1}\frac{%
s_{k}^{2}r^{2}+\left( 4s_{k}^{2}-3\right) z^{2}}{(s_{k}^{2}r^{2}+z^{2})^{3}}.
\label{UbrizLargeinq}
\end{equation}%
Assuming $z\gg r$, for $q\geqslant 2$, we find from
(\ref{UbrizLargeinq}) that up to the leading order
$U_{b}(\mathbf{r})\approx -q\alpha (0)/(8\pi z^{4})$. For a plate
in Minkowski spacetime the corresponding asymptotic is given by
the formula $U_{b}^{\mathrm{(M)}}(\mathbf{r})\approx -3\alpha
(0)/(8\pi z^{4})$.

For the isotropic polarizability and at distances smaller than the
wavelength of the dominant atomic transitions wavelength the asymptotic of
the CP potential has the form%
\begin{eqnarray}
U_{b}(\mathbf{r}) &\approx &-\frac{1}{4\pi z^{3}}\int_{0}^{\infty
}du\,\alpha (iu)\bigg[\sideset{}{'}{\sum}_{k=0}^{[q/2]}g_{2}(r/z,s_{k})
\notag \\
&&-\frac{q}{\pi }\int_{0}^{\infty }dy\frac{\sin (q\pi )g_{2}(r/z,\cosh y)}{%
\cosh (2qy)-\cos (q\pi )}\bigg],  \label{Ubizsm}
\end{eqnarray}%
where%
\begin{equation}
g_{2}(y,x)=(x^{2}-1)\frac{y^{2}x^{2}-2}{(y^{2}x^{2}+1)^{5/2}}.  \label{g2y}
\end{equation}%
For integer values $q$ this gives%
\begin{equation}
U_{b}(\mathbf{r})\approx \frac{1}{8\pi }\sum_{k=0}^{q-1}(1-s_{k}^{2})\frac{%
r^{2}s_{k}^{2}-2z^{2}}{(r^{2}s_{k}^{2}+z^{2})^{5/2}}\int_{0}^{\infty
}du\,\alpha (iu).  \label{Ubizint}
\end{equation}%
In particular, for $r\ll z$ and $q\geqslant 2$, from (\ref{Ubizint}), we
find $U_{b}(\mathbf{r})\approx -q\int_{0}^{\infty }du\,\alpha (iu)/(8\pi
z^{3})$. For the CP potential in the Minkowski spacetime at small distances
from the conducting plate one has $U_{b}^{\mathrm{(M)}}(\mathbf{r})\approx
-\int_{0}^{\infty }du\,\alpha (iu)/(4\pi z^{3})$.

\section{Oscillator model}

\label{Sec:Osc}

For further transformation of the general formula (\ref{Ub1}), the frequency
dependence of the polarizability tensor appearing in (\ref{hlp}) should be
specified. For the eigenvalues of the polarizability tensor we use the
anisotropic oscillator model. In this model,%
\begin{equation}
\alpha _{n}(i\xi )=\sum_{j}\frac{g_{j}^{(n)}}{\omega _{j}^{(n)2}+\xi ^{2}},
\label{alfll}
\end{equation}%
where $\omega _{j}^{(n)}$ and $g_{j}^{(n)}$ are the oscillator frequencies
and strengths, respectively. For the functions (\ref{hlp}) we find the
expressions%
\begin{eqnarray}
h_{lp}(y) &=&y^{2}\sum_{n=1}^{3}\beta
_{ln}^{2}\sum_{j}g_{j}^{(n)}B_{p}(y\omega _{j}^{(n)}),  \notag \\
h(y) &=&y^{2}\sum_{n=1}^{3}\beta _{1n}\beta
_{3n}\sum_{j}g_{j}^{(n)}\sum_{p=1}^{3}h_{p}B_{p}(y\omega _{j}^{(n)}),
\label{hlposc}
\end{eqnarray}%
with $h_{1}=h_{2}=3$, $h_{3}=1$, and%
\begin{equation}
B_{p}(x)=\int_{0}^{\infty }du\frac{u^{p-1}e^{-u}}{u^{2}+x^{2}}.  \label{Bp}
\end{equation}%
For the first two functions in (\ref{Bp}) one has
\begin{eqnarray}
B_{1}(x) &=&x^{-1}\left[ \sin (x)\text{Ci}(x)-\cos (x)\text{si}(x)\right] ,
\notag \\
B_{2}(x) &=&-\cos (x)\text{Ci}(x)-\sin (x)\text{si}(x),  \label{Bp12}
\end{eqnarray}%
where the functions Ci$(x)$ and si$(x)$ are defined in \cite{Abra72}. The
functions $B_{p}(x)$ for $p\geqslant 3$ are obtained by using the recurrence
formula%
\begin{equation}
B_{p+2}(x)=\Gamma (p)-x^{2}B_{p}(x).  \label{Bprec}
\end{equation}

Now the expression for the CP potential is given by (\ref{Ub1}) where%
\begin{eqnarray}
&& f(r,z,x) =4\sum_{n=1}^{3}\sum_{j}g_{j}^{(n)}\sum_{p=1}^{3}\frac{%
B_{p}(2\omega _{j}^{(n)}\sqrt{r^{2}x^{2}+z^{2}})}{(r^{2}x^{2}+z^{2})^{2}}
\notag \\
&& \qquad \times \left\{ \sum_{l=1}^{3}\left[
(r^{2}x^{2}+z^{2})b_{lp}(x)+z^{2}c_{lp}(x)\right] \beta
_{ln}^{2}+2h_{p}rzx^{2}\beta _{1n}\beta _{3n}\right\} .
\label{frzOsc}
\end{eqnarray}%
Note that for a plate in the Minkowski spacetime the CP potential is given
by the expression%
\begin{equation}
U_{b}^{\mathrm{(M)}}(\mathbf{r})=-\frac{1}{8\pi z^{2}}\sum_{n=1}^{3}%
\sum_{j}g_{j}^{(n)}\left\{ [B_{1}(2\omega _{j}^{(n)}z)+B_{2}(2\omega
_{j}^{(n)}z)]\left( 1+\beta _{3n}^{2}\right) +B_{3}(2\omega
_{j}^{(n)}z)\left( 1-\beta _{3n}^{2}\right) \right\} .  \label{UbMOsc}
\end{equation}%
This potential is a monotonic increasing function of $z$ and the
corresponding force is attractive for all distances.

At small distances, $\omega _{j}^{(n)}\sqrt{r^{2}+z^{2}}\ll 1$, the dominant
contribution comes from the term with $p=1$ by using the asymptotic
expression $B_{1}(y)\approx \pi /(2y)$, valid for $y\ll 1$. If in addition $%
r\ll z$ one finds%
\begin{equation}
f(r,z,x)\approx \frac{\pi }{z^{3}}\sum_{n=1}^{3}\sum_{j}\frac{g_{j}^{(n)}}{%
\omega _{j}^{(n)}}\left[ 1+\beta _{3n}^{2}-2x^{2}\left( 1-\beta
_{3n}^{2}\right) \right] .  \label{frzOsc2}
\end{equation}%
For integer $q\geqslant 2$, by using this expression, for the plate-induced
part in the CP potential we obtain the asymptotic expression
\begin{equation}
U_{b}(\mathbf{r})\approx -\frac{q}{16z^{3}}\sum_{n=1}^{3}\sum_{j}\frac{%
g_{j}^{(n)}}{\omega _{j}^{(n)}}\beta _{3n}^{2}.  \label{UbOscsm}
\end{equation}%
For a plate in Minkowski spacetime, at distances smaller than the wavelength
of the main atomic absorption lines, in the leading order one has
\begin{equation}
U_{b}^{\mathrm{(M)}}(\mathbf{r})\approx -\frac{z^{-3}}{32}%
\sum_{n=1}^{3}\sum_{j}\frac{g_{j}^{(n)}}{\omega _{j}^{(n)}}\left( 1+\beta
_{3n}^{2}\right) .  \label{UbMsmall}
\end{equation}%
In the opposite limit of large distances, $\omega _{j}^{(n)}\sqrt{r^{2}+z^{2}%
}\gg 1$, we use $B_{p}(z)\approx \Gamma (p)/z^{2}$, $z\gg 1$ and the result (%
\ref{flarge}) is recovered with $\alpha _{n}(0)=\sum_{j}g_{j}^{(n)}/\omega
_{j}^{(n)2}$.

In the isotropic case $g_{j}^{(n)}=g_{j}$, $\omega _{j}^{(n)}=\omega _{j}$
and the expression (\ref{frzOsc}) reduces to%
\begin{eqnarray}
f(r,z,x) &=&8\sum_{j}g_{j}\left\{ B_{3}(y_{j})\frac{%
z^{2}(1-2x^{2})-r^{2}x^{4}}{(r^{2}x^{2}+z^{2})^{2}}\right.  \notag \\
&&\left. +(x^{2}-1)\left[ B_{1}(y_{j})+B_{2}(y_{j})\right] \frac{%
r^{2}x^{2}-2z^{2}}{(r^{2}x^{2}+z^{2})^{2}}\right\} ,  \label{frzOscIz2}
\end{eqnarray}%
with the notation%
\begin{equation}
y_{j}=2\omega _{j}\sqrt{r^{2}x^{2}+z^{2}}.  \label{yj}
\end{equation}%
For $y_{j}\gg 1$ we obtain the result (\ref{Ubrizlarge}) with $\alpha
_{n}(0)=\sum_{j}g_{j}\omega _{j}^{-2}$. An asymptotic expression for the CP
potential at small distances, corresponding to $y_{j}\ll 1$, is obtained
from (\ref{Ubizsm}) with the substitution $\int_{0}^{\infty }du\,\alpha
(iu)=(\pi /2)\sum_{j}g_{j}/\omega _{j}$.

Note that in the isotropic case for the pure string part one has \cite%
{Saha11EPJ}%
\begin{equation}
U_{0}(\mathbf{r})=\frac{1}{2\pi }\left[ \sum_{k=1}^{[q/2]}f_{0}(r,s_{k})-%
\frac{q}{\pi }\int_{0}^{\infty }dy\frac{\sin (q\pi )f_{0}(r,\cosh y)}{\cosh
(2qy)-\cos (q\pi )}\right] ,  \label{U0OscIz}
\end{equation}%
where%
\begin{equation}
f_{0}(r,v)=\sum_{j}\frac{g_{j}}{r^{2}v^{2}}\left\{ v^{2}\left[
B_{1}(2rv\omega _{j})+B_{2}(2rv\omega _{j})\right] +\left( 1-v^{2}\right)
B_{3}(2rv\omega _{j})\right\} .  \label{f0}
\end{equation}

\begin{figure}[tbph]
\begin{center}
\epsfig{figure=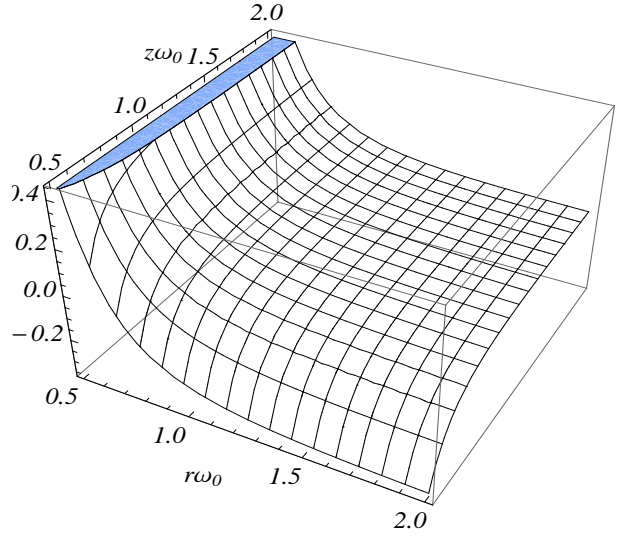,width=8.cm,height=7.cm}
\end{center}
\caption{CP potential, $U(\mathbf{r})/(g_{0}\protect\omega _{0}^{2})$, as a
function of the rescaled distances from the wall and from the string in the
conical space with the parameter $q=3$.}
\label{fig2}
\end{figure}

As a numerical example, in figure \ref{fig2} we plot the dependence of the
CP potential of the microparticle with an isotropic polarizability tensor on
the distances from the wall and from the string. Single-oscillator model is
used for the polarizability. For the parameter $q$ describing the conical
space we have taken the value $q=3$. As it is seen from the plot, the
Casimir-Polder force is repulsive with respect to the string and attractive
with respect to the wall.

As we have mentioned before, for an anisotropic polarizability tensor the CP
potential, in addition to the coordinates $r$ and $z$ of the polarizable
particle, depends also on the orientation of the principal axes for the
polarizability tensor. As a result of this dependence a moment of force acts
on the microparticle. In the numerical example below we use the single
oscillator model with $\alpha _{n}(i\xi )=g^{(n)}/[\omega ^{(n)2}+\xi ^{2}]$
and $g^{(1)}=g^{(2)}$, $\omega ^{(1)}=\omega ^{(2)}$. In this case the CP
potential depends on the angles $\alpha $ and $\beta $ only. In figure \ref%
{fig3} we display the dependence of the CP potential, $r^{2}U(\mathbf{r}%
)/g^{(1)}$, as a function of the angles $\alpha $ and $\beta $ for $q=3$, $%
\omega ^{(1)}r=1$, $\omega ^{(1)}z=1$, $\omega ^{(3)}/\omega ^{(1)}=1.5$,
and $g^{(3)}/g^{(1)}=1.25$. The values for the angles $\alpha $ and $\beta $
corresponding to the minimum of the potential determine the equilibrium
orientation of the principal axes for the polarizability tensor.

\begin{figure}[tbph]
\begin{center}
\epsfig{figure=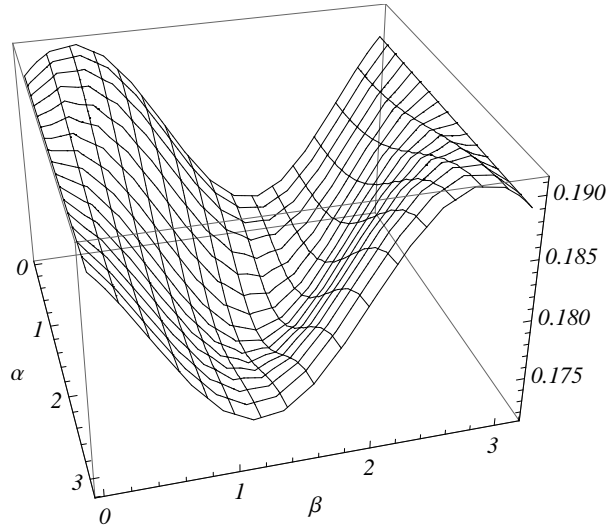,width=8.cm,height=7.cm}
\end{center}
\caption{CP potential, $r^{2}U(\mathbf{r})/g^{(1)}$, as a function of the
angles $\protect\alpha $ and $\protect\beta $ for an anisotropic
polarizability tensor. The values of the parameters used in the numerical
evaluation are given in the text.}
\label{fig3}
\end{figure}

\section{Conclusions}

\label{Sec:Conc}

In this paper we have investigated the CP interaction between a polarizable
microparticle and a conducting plate in a conical space. The corresponding
potential is expressed in terms of the retarded Green tensor for the
electromagnetic field by formula (\ref{eq2}). This tensor contains the
information about the physical and geometrical properties of vacuum
fluctuations. For the evaluation of the Green tensor we have used the direct
mode-summation method. In this way the Green tensor is decomposed into the
boundary-free and plate-induced parts. The CP interaction in the
boundary-free conical space has been discussed previously and here we were
mainly concerned with the effects related to the presence of the conducting
plate. The corresponding contribution to the Green tensor is given by the
expressions (\ref{Gbll}) and (\ref{G13bc}) for the diagonal and off-diagonal
components, respectively.

Similarly to the Green tensor, the CP\ potential is decomposed as (\ref{Udec}%
) with the boundary-induced part given by the expression (\ref{Ub1}). The
corresponding force depends on the distance of the microparticle from the
string, on the distance from the plate and on the orientation of the
polarizability tensor principal axes. The dependence on the orientation
enters into the potential through the dependence of the components of the
polarizability tensor on the Euler angles. The latter is given by the
formulas (\ref{alfllom}) and (\ref{alf13}) with the matrix $\hat{R}$ given
by (\ref{betmatrix}). With dependence of the polarizability tensor
eigenvalues and the orientation of the principal axes, the CP force can be
either attractive or repulsive. The general formula is simplified in the
special case with integer values of the parameter $q=2\pi /\phi _{0}$ [see (%
\ref{Ubqint})]. At distances much larger than the relevant transition
wavelengths, the expression for the function $f(r,z,x)$ appearing in the
expression for the CP potential takes the form (\ref{flarge}). If in
addition $z\gg r$, one has the asymptotic (\ref{frzAs}). In this case the
potential varies inversely with the fourth power of the distance from the
conducting plate and the corresponding force is attractive with respect to
the plate. For integer values of $q\geqslant 2$, the asymptotic behavior of
the potential is given by the expression (\ref{UbAsInt}) which depends on
the orientation of the polarizability tensor principal axes. For a plate in
Minkowski spacetime the corresponding asymptotic expression is given by (\ref%
{UbmAs}) and in the leading order the force does not depend on the
orientation. For the isotropic polarizability tensor the plate-induced part
in the CP potential is given by the expression (\ref{Ub1}) with the function
$f(r,z,x)$ given by (\ref{frzxiz}). At large distances and for integer
values of $q$, the corresponding asymptotic expression is given by (\ref%
{UbrizLargeinq}). If in addition $z\gg r$, for $q\geqslant 2$ one has $%
U_{b}/U_{b}^{\mathrm{(M)}}\approx q/3$.

It is important to call attention to the fact that $U_{b}$ is divergent at $%
z=0$. Otherwise, near the string, its behavior is well defined. For the
frequency dependence of the polarizability tensor we have used the
anisotropic oscillator model with the eigenvalues given by (\ref{alfll}).
With this model, the function $f(r,z,x)$ in the expression (\ref{Ub1}) for
the CP\ potential takes the form (\ref{frzOsc}) for the general case and the
form (\ref{frzOscIz2}) in the case of isotropic polarizability tensor. In
the case of anisotropic polarizability, the dependence of the CP potential
on the orientation of the polarizability tensor principal axes also leads to
a moment of force acting on the particle. This results in the macroscopic
polarization of a system of particles induced by combined effects of the
string and the boundary.

In the discussion above we have assumed that the electromagnetic field is
prepared in the vacuum state. If the field is prepared in a thermal state
with temperature $T$ a new length scale appears, $\lambda _{T}=(k_{B}T)^{-1}$%
, with $k_{B}$ being the Boltzmann constant. At nonzero temperature the
results obtained in this paper remain valid in the region $r,z\ll \lambda
_{T}$.

\section*{Acknowledgments}

E.R.B.M., V.B.B. and H.F.M. thank Conselho Nacional de Desenvolvimento Cient%
\'{\i}fico e Tecnol\'{o}gico (CNPq) for partial financial support.

\end{document}